\documentclass[prd,%twocolumn,
showpacs,floatfix]{revtex4}
\usepackage{bm}
\usepackage{amssymb}
\usepackage{graphicx}
\usepackage{epstopdf}

\begin{document}
\title{Energy in one dimensional linear waves in a string}
\author{Lior M. Burko}
\affiliation{
Department of Physics and Center for Space Plasma and Aeronomic Research, University of Alabama in Huntsville, Huntsville, Alabama 35899, USA}
\date{July 22, 2010}
\begin{abstract}
We consider the energy density and energy transfer in small amplitude, one--dimensional waves on a string, and find that the common expressions used in textbooks for the introductory physics with calculus course give wrong results for some cases, including standing waves. We discuss the origin of the problem, and how it can be corrected in a way appropriate for the introductory calculus based physics course. 
\end{abstract}
\maketitle

%\documentclass[prd,%twocolumn,
%floatfix,nofootinbib]{revtex4}
%\usepackage{bm}ci
%\usepackage{amssymb}
%\usepackage{graphicx}
%\usepackage{epstopdf}

%\begin{document}

%\title{Energy in one dimensional linear waves in a string}

%\author{Lior M.~Burko}

%\affiliation{Department of Physics and Center for Space Physics and Aeronomic Research, University of Alabama in Huntsville, Huntsville, Alabama 38599, USA}

%\date{\today}

%\begin{abstract}
%We consider the energy density and energy transfer in small amplitude, one--dimensional waves on a string, and find that the common expressions used in textbooks for the introductory physics with calculus course give wrong results for some cases, including standing waves. We discuss the origin of the problem, and how it can be corrected in a way appropriate for the introductory calculus based course. 
%\end{abstract}
%\pacs{01.55.+b}

%\maketitle

\section{Introduction}

Introductory textbooks say that the potential energy density of linear tension waves in a one--dimensional string is given by 
$\,du=(1/2)\tau\,(\,\partial y/\,\partial x)^2$, where $y(t,x)$ is the displacement from equilibrium, $\tau=\mu v^2$ is the string tension, $\mu$ is the linear mass density and $v$ the wave speed.  The kinetic energy density is given by $\,dk=(1/2)\mu (\,\partial y / \,\partial t)^2$, such that the total mechanical energy density is simply, 
\begin{equation}\label{eq1}
\,d\varepsilon = \,dk+\,du=
\frac{1}{2}\,\mu\left[\left(\frac{\,\partial y}{\,\partial t}\right)^2+v^2\, \left(\frac{\,\partial y}{\,\partial x}\right)^2\right]\, .
\end{equation}
Specifically, introductory textbooks say that the elastic potential energy is quadratic in the slope of the string \cite{catI}. Recalling that the wave's speed $v=\,dx/\,dt$, we see that the potential and kinetic energy densities are equal. 

We have found that most introductory physics with calculus textbooks can be arranged in four main categories:  Category I books include a discussion of the potential energy being proportional to the square of the spatial derivative of the displacement, similarly to Eq.~(\ref{eq1}) (see also below) \cite{catI}. Category II books discuss the energy in string waves based on an analogy with a harmonic oscillator, either a spring or a pendulum \cite{catII}. Category III books discuss the power in the waves, but not the potential energy density itself, and find the power calculating the work done by the elastic force \cite{catIII}. Finally, Category IV books ignore the question altogether \cite{catIV}.

In view of Eq.~(\ref{eq1}), in a traveling wave the potential and kinetic energies are in phase: they are both maximal together, and they vanish together too. Specifically, at a crest or a trough the kinetic energy vanishes as the string element is momentarily stationary, and the potential energy of the string element vanishes because the string element is horizontal. At zero displacement both are maximal, as both the string transverse velocity and the string's slope are maximal. The total mechanical energy of the string element therefore oscillates, as indeed should be expected as energy is transported with the wave along the string. 

Consider now a standing wave. At the antinodes of the wave the string element is always horizontal, and therefore one may expect, based on the statements made in introductory textbooks, that there is no potential energy. The kinetic energy of the same string element is zero at maximum displacement, and is maximal at zero displacement. Therefore, the total mechanical energy of  the string element is not constant: it oscillates between a maximal value at zero displacement and zero value at maximal displacement.  This conclusion presents us with an immediate conundrum: where does the energy go? Unlike the case of a traveling wave, in a standing wave there is no energy transfer, and the total mechanical energy of each string element is expected to be stationary. The same problem comes about when considering the wave's nodes: As the nodal points are never in motion, the kinetic energy always vanishes. The slope of the string, however, oscillates between zero (when the wave is at zero displacement) and a maximal value (when the wave is at maximal displacement). As in the case of the antinodes, we are presented with the same conundrum: What happens to the energy in a standing wave? and how come the energy of a string segment is time dependent, when clearly there cannot be any transport of energy along the string? These are fundamental questions whose discussion  may benefit the conceptual understanding of students of the introductory physics with calculus course. Surprisingly, not even a single textbook we have surveyed includes any discussion of these questions. 

In this paper we consider the question of the energy content and energy transfer in string waves,  point out why the description of the potential energy in terms of the slope of the string fails for standing waves, and how to treat the potential energy in a way that is applicable to both traveling and standing waves. We assume throughout that the string has zero potential energy at equilibrium, that the string is frictionless, and that no energy is supplied to drive the waves. We further assume that the tension $\tau$ is uniform along the string, which is equivalent to idealizing the string as having vanishing Young's modulus, so that no nonlinear effects excite longitudinal modes in the string. This paper is particularly relevant to physics and engineering majors who take the introductory physics with calculus course, and to instructors of such courses. 

\section{Common derivation of the string's energy}

The potential energy associated with elastic waves in a string can be found using  a number of different derivations. The derivations based on the virtual displacement method is typically inappropriate for the introductory course, because of both the use of the concept of virtual displacement, that most students encounter for the first time in an intermediate mechanics course, and because it involves sophisticated arguments on why a surface term is negligible \cite{mathews}. Other methods are the derivation from the stress and strain in the string, and energy conservation considerations \cite{mathews}. 

An elementary derivation that appears in a number of introductory textbooks is as follows \cite{catI}: Let an element of the unstretched string (in equilibrium) be $\,dx$, and the same element when displaced from equilibrium, is of length $\,ds$. Therefore, the lengthening of the string is by $\,ds-\,dx=\sqrt{\,dx^2+\,dy^2}-\,dx\approx \left[1+\,(1/2)(\,\partial y/\,\partial x)^2+\cdots\right]\,dx-\,dx\approx(1/2)(\,\partial y/\,\partial x)^2\,dx$. The potential energy in the string is related to the restoring force by $F=\,dU/(\,ds-\,dx)$. The restoring force $F$ is just the string tension $\tau$. Therefore, 
$$\,dU=\tau (\,ds-\,dx)=\frac{1}{2}\tau \left(\frac{\,\partial y}{\,\partial x}\right)^2\,dx\, , $$
and the potential energy density is 
\begin{equation}\label{pot_en}
\,du=\frac{1}{2}\tau \left(\frac{\,\partial y}{\,\partial x}\right)^2\, . 
\end{equation}

We can write the total power as
\begin{eqnarray*}
\frac{\,dE}{\,dt}&=&\frac{\,dK}{\,dt}+\frac{\,dU}{\,dt}\\
&=&\frac{1}{2}\mu v\,\left(\frac{\,\partial y}{\,\partial t}\right)^2+\frac{1}{2}v\tau\,\left(\frac{\,\partial y}{\,\partial x}\right)^2
\end{eqnarray*}
or
\begin{eqnarray}\label{e1}
\,dE_1=\frac{1}{2}\mu \left[\,\left(\frac{\,\partial y}{\,\partial t}\right)^2+v^2\,\left(\frac{\,\partial y}{\,\partial x}\right)^2\right]\, dx
\end{eqnarray}
using $\tau=\mu v^2$. As $v=\,dx/\,dt$, it is clear that there is as much power in the kinetic and potential parts, so that the momentary values
$\frac{\,dK}{\,dt}=\frac{\,dU}{dt}$. Equation (\ref{e1}) is the common expression appearing in introductory textbooks for the total energy in the string element $\,dx$, which we denote by $\,dE_1$ \cite{catI}. Before we confront it with an alternative expression, let us list some of the properties of this expression.  
The kinetic and potential energies are equal. This means that when the material element of the string is at a crest or a trough, there is zero energy in the wave in both kinetic and potential energies which vanish. When the element is going through equilibrium, both kinetic and potential energies are at maximum (notice that the slope of the wave function is maximal when the material element is passing through the equilibrium state). This is how energy can be transferred in a traveling wave: When the material element is at the equilibrium point there is maximum energy in the string. A quarter period later, that material element is at a crest or a trough, and there is zero energy in it. That energy has propagated along the string to a material element that had zero energy before, and so on. We can have another interpretation of this result by noticing that we can write the radiated power by the string element as $\,dP=\,d{\bf F}\cdot{\bf u}$, where $u:=\,\partial y/\,\partial t$ is the transverse string velocity (not the wave speed!), and $\,dF=\tau\,\left(\frac{\,\partial^2 y}{\,\partial x^2}\right)\,dx$ is the component of the net force acting on the string element (i.e., the difference in the force between the two endpoints of the element) in the direction of the element's motion. The power therefore equals the work done by the restoring force, as one should indeed expect from conservation of energy. As we show below, this expression is problematic for the description of energy in a standing wave.

\section{String's energy from work done to stretch the string to momentary shape}

We follow here the derivation in \cite{mf}. We describe the shape of the string by $y(t,x)$. The potential energy associated with this shape of the string is associated with the work done to stretch the string from its equilibrium shape $y=0$ to $y(t,x)$. At each point in space (each value of $x$) we introduce a parameter $\beta\in [0,1]$, such that $\beta=0$ corresponds to the equilibrium configuration, and $\beta=1$ corresponds to $y(t,x)$. The transverse force on any string element of length $\,dx$ at the intermediate point $\beta$ is given by $f_{\beta y}=\tau\,\partial^2 (\beta y)/\,\partial x^2\, dx=\beta\tau\,\partial^2 y/\,\partial x^2\, dx$. As a note made in passing, we comment that the wave equation for the string is conventionally obtained when this expression is equated with $\mu\,dx (\,\partial^2 (\beta y)/\,\partial t^2)\, .$ The work done moving the element from $\beta$ to $\beta +\,d\beta$ is therefore $f_{\beta y}\,d(\beta y)=f_{\beta y}y\,d\beta = y \tau (\,\partial^2 y/\,\partial x^2)\, dx\beta \,d\beta$. The total work required to get this string element to the actual configuration $y(t,x)$ is therefore
$\,dW= \tau y\,(\,\partial^2 y/\,\partial x^2)\, dx\int_0^1\beta \,d\beta=\frac{1}{2} \tau y\, (\,\partial^2 y/\,\partial x^2)\, dx$. The change in potential energy equals the negative of the work, so that 
$$\,dU=-\frac{1}{2} \tau\, y\, \left(\frac{\,\partial^2 y}{\,\partial x^2}\right)\, dx\, .$$
The total energy is therefore
\begin{eqnarray*}
\,dE=\,dK+\,dU&=&\frac{1}{2}\mu\, \left(\frac{\,\partial y}{\,\partial t}\right)^2-\frac{1}{2} \tau\, y\, \left(\frac{\,\partial^2 y}{\,\partial x^2}\right)\, dx\end{eqnarray*}
\begin{eqnarray}\label{e2}
\,dE_2=
\frac{1}{2}\mu\left[\, \left(\frac{\,\partial y}{\,\partial t}\right)^2-v^2\,y\, \left(\frac{\,\partial^2 y}{\,\partial x^2}\right)\right]\,dx
\end{eqnarray}
Notice that the potential energy term $\,dU>0$, as for any simple oscillatory function ${\rm sgn} (y)\times{\rm sgn} (\,\partial^2 y/\,\partial x^2)=-1$. 

\subsection*{Comparison of the two expressions}

The two expressions we have found for the total energy of the string element $\,dx$ ---Eqs.~(\ref{e1}) and (\ref{e2})--- are unequal. First, we show how the two are related. Then, we discuss the meaning of the different results for $\,dE_1$ and $\,dE_2$. The two expressions are locally equal (i.e., equal for the differential element $\,dx$) if $y'^2=-yy''$, where a prime denotes differentiation with respect to $x$. Only a very particular function, specifically $y(t,x)=\pm\sqrt{a(t)\,x+b(t)}$ where $a,b$ are arbitrary functions of the time $t$, satisfies this relation. For any other wave profile the two are indeed unequal. 

We next examine the global agreement of the two expression, $\,dE_1$ and $\,dE_2$, i.e., when the two agree for the entire string, or at least for a finite length of the string. 
$$\left. E_1\right|_{x_1}^{x_2}=\int_{x_1}^{x_2}\,dE_1=\frac{1}{2}\mu\,\int_{x_1}^{x_2} \left[\,\left(\frac{\,\partial y}{\,\partial t}\right)^2+v^2\,\left(\frac{\,\partial y}{\,\partial x}\right)^2\right]\, dx$$
assuming a homogeneous string. To find $\left. E_2\right|_{x_1}^{x_2}=\int_{x_1}^{x_2}\,dE_2$ we integrate by parts, 
and find that 
$$\left. E_2\right|_{x_1}^{x_2}=\frac{1}{2}\mu\,\int_{x_1}^{x_2} \left[\,\left(\frac{\,\partial y}{\,\partial t}\right)^2+v^2\,\left(\frac{\,\partial y}{\,\partial x}\right)^2\right]\, dx-\frac{1}{2}\tau\left[y\,\left(\frac{\,\partial y}{\,\partial x}\right)\right]_{x_1}^{x_2}\, .$$

%so that 
%\begin{eqnarray*}
%\left. E_2\right|_{x_1}^{x_2}=\int_{x_1}^{x_2}\,dE_2&=&\frac{1}{2}\mu\,\int_{x_1}^{x_2} \left[\,\left(\frac{\,\partial y}{\,\partial t}\right)^2-v^2\,y\,\left(\frac{\,\partial^2 y}{\,\partial x^2}\right)\right]\, dx\\
%&=&\frac{1}{2}\mu\,\int_{x_1}^{x_2} \left[\,\left(\frac{\,\partial y}{\,\partial t}\right)^2+v^2\,\left(\frac{\,\partial y}{\,\partial x}\right)^2\right]\,dx-\frac{1}{2}\tau\left[y\,\left(\frac{\,\partial y}{\,\partial x}\right)\right]_{x_1}^{x_2}\\
%&=&\left. E_1\right|_{x_1}^{x_2}-\frac{1}{2}\tau\left[y\,\left(\frac{\,\partial y}{\,\partial x}\right)\right]_{x_1}^{x_2}\, .
%\end{eqnarray*}

The difference between the two results, 
\begin{equation}\label{diff}
\left. E_2\right|_{x_1}^{x_2}-\left. E_1\right|_{x_1}^{x_2} = -\frac{1}{2}\tau\left[y\,\left(\frac{\,\partial y}{\,\partial x}\right)\right]_{x_1}^{x_2}\, ,
\end{equation}
is a boundary term, reflecting the energy at the two end points of the string (or the end points of the finite segment in question). This result appears in \cite{mf}. When the string is infinitely long, the boundary term vanishes, because any finite disturbance will not have had enough time to arrive at infinity. When the string is of finite length, with either fixed or free end point boundary conditions, again the boundary term vanishes, and similarly also for periodic boundary conditions. We therefore see that the global energy content of the string is the same whether we calculate $\left. E_1\right|_{x_1}^{x_2}$ or $\left. E_2\right|_{x_1}^{x_2}$ under very broad conditions, even though $\,dE_1\ne\,dE_2$.

\section{Application of results to a standing wave}

To show the local disagreement of Eq.~(\ref{e1}) and Eq.~(\ref{e2}), and how the former leads to nonsensical results in some cases while the latter leads to correct results, consider a standing wave, and specifically take the wave function to be $y(t,x)=y_0\,\sin(kx)\,\sin(\omega t)$. First, we find $\,dE_1$, 
$$\,dE_{1}=\frac{1}{2}\mu \left[\,\left(\frac{\,\partial y}{\,\partial t}\right)^2+v^2\,\left(\frac{\,\partial y}{\,\partial x}\right)^2\right]_x\, dx
=\frac{1}{4}\mu\omega^2y_0^2\left[1-\cos(2kx)\,\cos(2\omega t)\right]\,dx\, ,$$
where the angular frequency  $\omega =kv$. This result for  $\,dE_1$ is explicitly time-dependent. This result, if accepted, would mean that the total energy of the string element oscillates with time with an amplitude that is position dependent, in direct violation of energy conservation: as this is a standing wave, there is no energy transfer, and each string element must conserve energy individually in the absence of friction. Specifically, at nodal points $kx_{\rm nodes}=n\pi$, so that 
$$\,dE_{1\;{\rm nodes}}=\frac{1}{2}\mu\,y_0^2\,\omega^2\,\sin^2\omega t\,dx\, ,$$ where the contributions are coming only from the potential energy term (the kinetic energy identically vanishes at nodes); at anti-nodes $kx_{\rm anti-nodes}=(2n+1)\pi/2$, so that 
$$\,dE_{1\;{\rm anti\;nodes}}=\frac{1}{2}\mu\,y_0^2\,\omega^2\,\cos^2\omega t\,dx\, , $$
where the contributions are coming only from the kinetic energy term (the potential energy term identically vanishes at anti-nodes).

%At the nodes the amplitude always vanishes, such that $kx_{\rm nodes}=n\pi$, when $n$ is an integer. Therefore, there is no kinetic energy. Using $\,dE_1$, 
%$$\,dE_{1\;{\rm nodes}}=\frac{1}{2}\mu \left[\,\left(\frac{\,\partial y}{\,\partial t}\right)^2+v^2\,\left(\frac{\,\partial y}{\,\partial x}\right)^2\right]_{\rm nodes}\, dx
%=\frac{1}{2}\mu\,y_0^2\,\omega^2\,\sin^2\omega t\,dx\, ,$$
%where the angular frequency  $\omega =kv$.
%This result, if accepted, would mean that the total energy of the string element oscillates with time, in direct violation of energy conservation: as this is a standing wave, there is no energy transfer, and each string element must maintain energy conservation individually in the absence of friction. At anti-nodes $kx_{\rm anti-nodes}=(2n+1)\pi/2$, so that the potential energy term vanishes, and only the kinetic energy contributes, such that 
%$$\,dE_{1\;{\rm anti\;nodes}}=\frac{1}{2}\mu \left[\,\left(\frac{\,\partial y}{\,\partial t}\right)^2+v^2\,\left(\frac{\,\partial y}{\,\partial x}\right)^2\right]_{\rm anti\;nodes}\, dx
%=\frac{1}{2}\mu\,y_0^2\,\omega^2\,\cos^2\omega t\,dx$$
%with the same problematic interpretation of energy conservation violation as in the case of nodes. 

Next, we find $\,dE_2$,  
$$\,dE_{2}=
\frac{1}{2}\mu\left[\, \left(\frac{\,\partial y}{\,\partial t}\right)^2-v^2\,y\, \left(\frac{\,\partial^2 y}{\,\partial x^2}\right)\right]_{x}\,dx=\frac{1}{2}\mu\,y_0^2\omega^2\,\sin^2 kx\,dx$$
which is explicitly time independent for any element of the string at all values of the time. In particular, at nodes 
$\,dE_{2\;{\rm nodes}}=0$, and at anti-nodes $\,dE_{2\;{\rm anti\;nodes}}=\frac{1}{2}\mu\,y_0^2\omega^2\,dx$.

%Using $\,dE_2$, the corresponding values for the energy of the string element at nodal or anti-nodal points are
%$$\,dE_{2\;{\rm nodes}}=
%\frac{1}{2}\mu\left[\, \left(\frac{\,\partial y}{\,\partial t}\right)^2-v^2\,y\, \left(\frac{\,\partial^2 y}{\,\partial x^2}\right)\right]_{\rm nodes}\,dx=0$$
%and
%$$\,dE_{2\;{\rm anti\;nodes}}=
%\frac{1}{2}\mu\left[\, \left(\frac{\,\partial y}{\,\partial t}\right)^2-v^2\,y\, \left(\frac{\,\partial^2 y}{\,\partial x^2}\right)\right]_{\rm anti\;nodes}\,dx=\frac{1}{2}\mu\,y_0^2\omega^2\,dx$$
%which are both time independent. In fact, we find the energy content of the string element $\,dx$ is time independent not just at nodes or anti-nodes, but rather anywhere along the string, as 
%$$\,dE_{2}=
%\frac{1}{2}\mu\left[\, \left(\frac{\,\partial y}{\,\partial t}\right)^2-v^2\,y\, \left(\frac{\,\partial^2 y}{\,\partial x^2}\right)\right]_{x}\,dx=\frac{1}{2}\mu\,y_0^2\omega^2\,\sin^2 kx\,dx$$
%and 

The total energy inside a wavelength is
$$\left.E_{2}\right|_{x=0}^{\lambda}
=\frac{1}{2}\frac{\mu}{k}\,y_0^2\omega^2\,\int_{kx=0}^{2\pi}\sin^2 kx\,d(kx)=\frac{\lambda}{4}\mu\,y_0^2\omega^2\, ,$$
where the wavelength $\lambda=2\pi/k$, 
which is the well known result. The same result is also found for $\left.E_{1}\right|_{x=0}^{\lambda}$:
\begin{eqnarray*}
\left.E_{1}\right|_{x=0}^{\lambda} &=& \frac{1}{4}\frac{\mu}{k}\,y_0^2\omega^2\,\int_{kx=0}^{2\pi}\left[1-\cos(2kx)\,\cos(2\omega t)
\right]\,d(kx)=\frac{\lambda}{4}\mu\,y_0^2\omega^2\\
&=&\left.E_{2}\right|_{x=0}^{\lambda} \, .
\end{eqnarray*}

The expressions for the energy in the differential string element $\,dx$ fail to agree because of the boundary term in Eq.~(\ref{diff}). Indeed, in the specific case considered here, this difference is
\begin{eqnarray*}
\left. E_2\right|_{x}^{x+\,dx}-\left. E_1\right|_{x}^{x+\,dx} &=& -\frac{1}{2}\tau\left[y\,\left(\frac{\,\partial y}{\,\partial x}\right)\right]_{x_1}^{x_2}\\
&=& \frac{1}{2}\,y_0^2\omega^2\,\left(\,\sin^2kx-\,\cos^2kx\,\right)\,dx=-\frac{1}{2}\,y_0^2\omega^2\,\cos(2kx)\, .
\end{eqnarray*}

To show how energy is exchanged between the kinetic and potential energies, let us consider an element of string $\,dx$, so that the amount of kinetic energy is
$\,dK=\frac{1}{2}\mu\,\omega^2y_0^2\,\sin^2(kx)\,\cos^2(\omega t)\,dx$ and the amount of potential energy is 
$\,dU=\frac{1}{2}\mu\,\omega^2y_0^2\,\sin^2(kx)\,\sin^2(\omega t)\,dx$. Their time rates of change are 
$\,d(\,dK)/\,dt=\mu\,\omega^3y_0^2\,\sin^2(kx)\,\sin(2\omega t)\,dx$ and 
$\,d(\,dU)/\,dt=-\mu\,\omega^3y_0^2\,\sin^2(kx)\,\sin(2\omega t)\,dx$ so that clearly 
$\,d(\,dE)/\,dt\equiv\,d(\,dK)/\,dt+\,d(\,dU)/\,dt=0$. 
In this case the {\em sum} of rates of change of kinetic and potential energies is zero. This implies that the energy of the string is constant. As there is no propagation of waves, there can be no transport of energy either. All that we have is sloshing of energy from kinetic to potential. 

Why does the elementary derivation for the potential energy as originating in the local stretching of a string element  fail? This derivation, leading to Eq.~(\ref{e1}) is based on the stretching amount of the local string element, without worrying what the two endpoints of the strings are doing. The differential energy in the element will therefore be potentially different form the actual energy by an amount that depends on the one--sided stretching of the edges. When we add together differential elements to find the energy content of a  finite string, the boundaries of neighboring elements cancel each other, so that the total difference is just on the boundaries of the finite string. When specially chosen boundary conditions are specified, we indeed find global agreement between the two expressions.

\section{Application of results to a traveling wave}

We revisit now the question of a traveling wave. Take $y(t,x)=y_0\,\sin(kx-\omega t)$. Using Eqs.~(\ref{e1}) and (\ref{e2}) to find the energy in the string element $\,dx$ we find
$$\,dE_1=\mu y_0^2\omega^2\,\cos^2(kx-\omega t)\,dx$$
and
$$\,dE_2=\frac{1}{2}\mu y_0^2\omega^2\,dx\, ,$$
respectively. The difference between the two expressions is stark: Eq.~(\ref{e2}) yields a constant energy for the string element $\,dx$ even though there is a traveling wave going through it, whereas Eq.~(\ref{e1}) yields an oscillatory result. The reason for this difference is that the two endpoints of the string element are not isolated, and there is flow of energy through them. Indeed, the difference between the two expressions is
$$\left.\,dE_2\right|_x^{x+dx}-\left.\,dE_1\right|_x^{x+dx}=\frac{1}{2}\mu y_0^2\omega^2\, \left[\,\sin^2(kx-\omega t)-\,\cos^2(kx-\omega t)\right]\,dx$$
to leading order in $\,dx$, which corresponds to the energy in the two end points of the string element. The interpretation of these expressions is that the total amount of energy in the string element is constant, but there is a time changing amount of energy in the end points. Neglecting the end points, there is therefore a time changing amount of energy in the element. The flow of energy through the end points corresponds to the transfer of energy along the string. 

\section*{Acknowledgments}

The author wishes to thank Seyed Sadeghi for discussions.  
This work has been supported in part by NASA/SSC grant No.~NNX07AL52A, and by NSF grants No.~PHY--0757344 and DUE--0941327.

\end{document}